\documentclass[twocolumn,nofootinbib,notitlepage]{revtex4-1}
\usepackage{epsfig,amsmath,slashed}
\usepackage{tensor}
\newcommand{\di}{{\rm d}}
\newcommand{\ii}{i}

\def\wO{{\widehat O}}
\def\wT{{\widehat T}}
\def\wj{{\widehat j}}
\def\wJ{{\widehat J}}

\def\wP{{\widehat P}}
\def\wA{{\widehat A}}
\def\wB{{\widehat B}}
\def\wC{{\widehat C}}
\def\wQ{{\widehat Q}}

\def\wrho{{\widehat{\rho}}}

\def\wTheta{\widehat{\Theta}}

\newcommand{\tr}{{\rm tr}}  
  
\newcommand{\e}{{\rm e}}

\newcommand{\be}{\begin{equation}}
\newcommand{\ee}{\end{equation}}                                                                               
\newcommand{\bea}{\begin{eqnarray}}
\newcommand{\eea}{\end{eqnarray}}                                                                               

\begin{document}
%*****************************************************************

%*****************************************************************************
\title{Quantum relativistic fluid at global thermodynamic equilibrium in curved spacetime} 
%*****************************************************************************

\author{F. Becattini, E. Grossi}\affiliation{Universit\`a di 
 Firenze and INFN Sezione di Firenze, Florence, Italy} 

\begin{abstract}
We present a new approach to the problem of the thermodynamical equilibrium of a 
quantum relativistic fluid in a curved spacetime in the limit of small curvature. 
We calculate the mean value of local operators by expanding the four-temperature 
Killing vector field in Riemann normal coordinates about the same spacetime point
and we derive corrections with respect to the flat spacetime expressions. Thereby, 
we clarify the origin of the terms proportional to Riemann and Ricci tensors introduced 
in general hydrodynamic expansion of the stress-energy tensor.
\end{abstract}

\maketitle

Thermal quantum field theory in flat spacetime is a well known subject. Its main goal 
is the calculation of mean values of local operators in the mixed state described 
by the grand-canonical density operator
\be\label{basic}
 \wrho = \frac{1}{Z} \exp[-\widehat H /T + \mu\widehat Q/T]
\ee
where $\widehat H$ is the hamiltonian, $T$ the temperature and $\mu$ the chemical
potential associated to a conserved charge $\wQ$. Its extension to curved spacetime 
features several difficulties, the main problem being the construction of a consistent 
theory of quantum fields in a non-Minkowskian background. Another problem is the 
construction of thermodynamical equilibrium states generalizing (\ref{basic})
\cite{solveen,pinamonti}. Indeed, for global equilibrium, a covariant expression 
of the eq.~(\ref{basic}) has been known for some time \cite{zuba,weert} 
\be\label{main}
 \wrho = \frac{1}{Z} \exp \left[ - \int_\Sigma \di \Sigma_\mu \left( 
 \wT^{\mu\nu} \beta_\nu - \zeta \wj^\mu \right) \right]
\ee
where $\wT$ is the stress-energy tensor built with quantum field operators, $\wj$
a conserved current, $\beta$ is the inverse temperature four-vector (shortly, 
four-temperature) field such that $T = 1/\sqrt{\beta^2}$ is the temperature measured 
by a (comoving) thermometer moving with velocity $u=\beta/\sqrt{\beta^2}$ and $\zeta=\mu/T$ 
\cite{becalocal}. As $\nabla_\mu \wT^{\mu\nu} = 0$, the eq.~(\ref{main}) is 
independent of the hypersurface $\Sigma$ and, therefore, $\wrho$ is stationary 
provided that $\zeta$ is constant and $\beta$ is a Killing vector field
\be\label{kill}
  \nabla_\mu \beta_\nu + \nabla_\nu \beta_\mu = 0
\ee
It should be emphasized that the direction of the Killing vector field $\beta$, 
that is $u$, can be interpreted as the velocity of the relativistic fluid whose 
stress-energy tensor operator is $\wT$ and that it {\em defines} a relativistic 
frame which does not necessarily coincide with the traditional Landau's or 
Eckart's \cite{becalocal}, as it was shown explicitely in ref.~\cite{becaquantum}. 

Suppose we want to calculate the (renormalized) mean value of a local operator $\wO$ 
at a point $x$ with the density operator (\ref{main}) 
\be\label{mean}
 \langle \wO(x) \rangle = \frac{1}{Z} \tr \left( \exp \left[ - \int_\Sigma \di \Sigma \;
  n_\mu ( \wT^{\mu\nu} \beta_\nu - \zeta \wj^\mu) \right] \wO(x) \right)_{\rm ren}
\ee
where $n$ is the unit normal vector of the hypersurface $\Sigma$.
In a curved spacetime, to make sense of (\ref{mean}) one should first make it precise 
the meaning of the operators, Hilbert spaces and traces, however we will see that 
in the limit of spacetimes whose curvature scale is much larger than thermal and 
microscopic interaction lengths, such issues can be avoided. 

If the field $\beta$ varies significantly over a much larger scale than the correlation 
length of $\wO$ and $\wT$, which is governed by the microscopic parameters of the theory 
(masses, couplings) and the proper temperature $T$, a Taylor expansion can be made 
about the point $x$ (we set $x=0$ without loss of generality)
\be\label{tayl}
 \beta_\nu(y) = \beta_\nu(0) + \partial_\lambda \beta_\nu(0) y^\lambda + 
 \frac{1}{2} \partial_\lambda \partial_\sigma \beta_\nu(0) y^\lambda y^\sigma 
 + \ldots
\ee
Using the above expansion into eq.~(\ref{mean}) and considering terms beyond the 
leading as small corrections, the mean value of the local operator $\wO(x)$ can be 
calculated by expanding the density operator with the Kubo identity, a well known 
method in linear response theoery. Thereby, $\langle \wO(x) \rangle$ is mainly 
determined by the four-temperature value $\beta$ in the same point $x$, that is 
(for $\zeta = 0$)
\be\label{reduced}
 \langle \wO (x) \rangle_\beta \simeq \frac{\tr \left( \exp \left[ - \beta_\nu(x)
 \int_\Sigma \di \Sigma(y) \; n_\mu(y) \wT^{\mu\nu}(y) \right] \wO(x) \right)} 
 {\tr \left( \exp \left[ - \beta_\nu(x) \int_\Sigma \di \Sigma(y) \; 
 n_\mu(y) \wT^{\mu\nu}(y) \right] \right)} 
\ee
plus corrections expressed by correlators such as
\be\label{correl}
  \int_\Sigma \di \Sigma(y) \; n_\mu (y) \left( \langle \wO(x) \wT^{\mu\nu}(y) 
  \rangle_\beta - \langle \wO(x) \rangle_\beta \langle \wT^{\mu\nu}(y) \rangle_\beta 
  \right)
\ee

The Taylor expansion of the four-temperature is the core of our method \cite{becalocal}. 
It differs from other approaches to relativistic hydrodynamics in curved spacetime 
based on the assumption of the existence of an extensive generating functional 
\cite{kaminski,tifr,japan}. We do not introduce such an assumption here and we do not 
even assume the existence of an entropy current. In fact, a simple expansion 
(\ref{tayl}) of the Killing vector generates integral terms in the eq.~(\ref{mean}) 
which are dependent on the choice of coordinates as well as on the hypersurface $\Sigma$, 
even though the full series will not. It would be then desirable to have a standard 
choice of both which makes the calculation of the individual terms of the expansion 
to have a direct physical meaning or, at least, to have a readily recognizable flat 
spacetime limit. In this respect, one would like to deal with curvilinear coordinates 
which are the closest approximation to Cartesian orthogonal coordinates at the point 
$x$ where the expansion is carried out. The natural choice is, thus, Riemann Normal 
Coordinates (RNC) \cite{rnc}; they are locally inertial ($\Gamma^\lambda_{\mu\nu}(x)=0$) 
and derivatives of the Christoffel symbols can be expressed as a simple linear 
combination of the Riemann tensor (we follow the conventions on metric and curvature 
tensors of ref.~\cite{landau})
\be\label{christ}
 \partial_\sigma \Gamma^\rho_{\lambda\nu}(x) = - \frac{1}{3} \left( 
  R\indices{^\rho_{\lambda\nu\sigma}}(x) + R\indices{^\rho_{\nu\lambda\sigma}}(x) 
  \right)
\ee
As to the hypersurface, $\Sigma$ can be conveniently chosen to be the closest approximation
to the cartesian 3-space of a comoving observer, namely the one which is normal to 
$\beta$ in the point $x=0$ and with the first Riemann normal coordinate $x^0 = 0$ 
fixed. Hence, the hypersurface $\Sigma$ is spanned by the three Riemann normal 
coordinates $x^1,x^2,x^3$, that is by three normal geodesics starting from the point 
$x=0$ in the three spacelike directions orthogonal to $\beta(0)$. Altogether, we 
can say that the calculation of integrals is carried out in the locally orthonormal 
frame of a free-falling observer with velocity $u= \beta/\sqrt{\beta^2}$ 
(that is, the $\beta$ frame \cite{becalocal}) and with a three-space described by the
three orthogonal spacelike geodesics. 

Finally, in a curved spacetime, the coefficients of a Taylor expansion of a vector 
field like in (\ref{tayl}) are not tensors as they are expressed by simple derivatives. 
This is a problem as it makes the truncated expression of the mean value (\ref{mean}) 
non-covariant starting from second order. To make it covariant, at least to second 
order, one has to expand, in inertial coordinates, the difference between two vector 
fields which coincide in the point $x=0$. It seems a natural choice to expand the 
difference between $\beta$ and $\bar\beta$ which results from the parallel transport 
of $\beta(0)$ along all the orthogonal outgoing geodesics from $x=0$. Therefore, we set
\be\label{expand}
 \int_\Sigma \di \Sigma \; n_\mu \wT^{\mu\nu} \beta_\nu =
 \int_\Sigma \di \Sigma \; n_\mu \wT^{\mu\nu} \left[ \bar\beta_\nu + (\beta_\nu
 -\bar\beta_\nu) \right]
\ee 
with $\beta(0)=\bar\beta(0)$, and expand the difference $\beta-\bar\beta$. 

At first order, in inertial coordinates, $\partial \bar\beta =0$ and, since the 
Killing equation (\ref{kill}) is a simple relation between partial derivatives, the 
gradient of $\beta$ field at $x=0$ is antisymmetric. Thus
\be\label{omega}
 \partial_\lambda (\beta_\nu - \bar\beta_\nu)(0) = \partial_\lambda \beta_\nu(0) 
  = \frac{1}{2} \left( \partial_\lambda \beta_\nu - \partial_\nu \beta_\lambda 
  \right)(0) \equiv \varpi_{\nu\lambda}(0)
\ee
which is defined as {\em thermal vorticity}. Since, in inertial coordinates in 
the point $x=0$
\bea\label{secder}
  && \nabla_\sigma\nabla_\lambda \beta_\nu = \partial_\sigma \partial_\lambda \beta_\nu
   - \partial_\sigma \Gamma^\rho_{\lambda\nu} \, \beta_\rho \nonumber \\
  && 0 = \partial_\sigma \partial_\lambda \bar\beta_\nu
   - \partial_\sigma \Gamma^\rho_{\lambda\nu} \, \bar\beta_\rho 
\eea
subtracting the two equations above we obtain
\be\label{secder2}
 \partial_\lambda \partial_\sigma(\beta_\nu - \bar\beta_\nu)(0) = 
 \nabla_\sigma \nabla_\lambda \beta_\nu(0) = R_{\rho\sigma\lambda\nu}(0) 
 \beta^\rho(0)
\ee
where the last equality is ought to $\beta$ being a Killing vector field \cite{weinberg}.
As the correlation length in the integral (\ref{correl}) is much smaller than curvature 
scales, to calculate the integrals one can also expand the integration measure 
$$ \di \Sigma = \di^3 y \; \sqrt{\gamma}$$ 
about the point $x=0$. The expansion of the 
induced metric $\gamma$ on the hypersurface $\Sigma$ at the second order in $y$ can be 
readily obtained by using the known features of RNC \cite{rnc}
\bea\label{measure}
 \sqrt{\gamma} \; && \simeq 1 - \frac{1}{6} (R_{\lambda\sigma}(0) - R_{0\lambda 0\sigma}(0))
  y^\lambda y^\sigma \nonumber \\
 && = 1 - \frac{1}{6} (R_{\lambda\sigma}(0) - R_{\rho\lambda\tau\sigma}(0)) u^\rho(0) u^\tau(0))
  y^\lambda y^\sigma 
\eea
where $R_{\lambda\sigma}$ is the Ricci tensor.

By using eqs.~(\ref{omega}),(\ref{secder2}),(\ref{measure}), the properties of the 
Riemann tensor indices, the expansion of the operator (\ref{expand}) at second order in 
the RNC can be written as
\bea\label{expand2}
 && \int_\Sigma \di \Sigma \; n_\mu \wT^{\mu\nu}(y) \beta_\nu 
 \simeq \int_\Sigma \di^3 y \; n_\mu \wT^{\mu\nu} \bar\beta_\nu \nonumber \\
 && - \frac{1}{2} \varpi_{\lambda\nu}(0) \int_\Sigma \di^3 y \; n_\mu 
 \left( y^\lambda \wT^{\mu\nu} - y^\nu \wT^{\mu\lambda} \right) \nonumber \\
 &&  + \frac{1}{2\sqrt{\beta^2}} B_{\lambda\sigma\nu}  \int_\Sigma \di^3 y \; n_\mu 
 y^\lambda y^\sigma \wT^{\mu\nu}
\eea
where $B$ is the adimensional tensor
\be\label{btensor}
  B_{\lambda\sigma\nu} = R_{\rho\lambda\sigma\nu} \sqrt{\beta^2} \beta^\rho - 
  \frac{1}{3} R_{\lambda\sigma} \sqrt{\beta^2} \beta_\nu - 
  \frac{1}{3} R_{\rho\lambda\sigma\tau} \beta^\rho \beta^\tau u_\nu
\ee

Because of the assumed large difference between curvature and microscopic scales,
the integration over $\Sigma$ can be approximated by an integration over its tangent 
timelike hyperplane $T\Sigma$ in $x=0$, identifying all parallel transported tensors  
in the in the eq.~(\ref{expand2}) with corresponding constant tensors in the tangent 
hyperplane. Particularly, for the hypersurface $\Sigma$ is spanned by spacelike 
geodesics, its normal vector $n$ is obtained by parallel transporting $n(0)$, which
is then to be identified with the time direction everywhere in the tangent spacetime; 
likewise, $\bar\beta$ is parallel transported by definition and it will have to 
correspond to the vector $\bar\beta = (1/T,{\bf 0})$ in the tangent spacetime. Indeed, 
the replacement of a curved with a flat spacetime makes it possible to recover all 
known features of quantum field theory. The operators can be interpreted as operators 
in the usual Hilbert spaces and - whenever applicable - their integrals mapped to 
the tangent hyperplane as generators of the Poincar\'e group of the tangent spacetime. So
\bea\label{generators}
 && \int_\Sigma \di^3 y \; n_\mu \wT^{\mu\nu}(y) \bar\beta_\nu(y) \rightarrow \bar\beta_\nu(0)
 \wP^\nu = \beta_\nu(0) \wP^\nu  \nonumber \\
 && \int_\Sigma \di^3 y \; n_\mu \left( y^\alpha \wT^{\mu\beta} - y^\beta \wT^{\mu\alpha} 
 \right) e_{\alpha}^{(\lambda)} e_{\beta}^{(\nu)} \rightarrow \wJ_0^{\lambda\nu}
\eea
where $\wP$ are four-momentum operators and $\wJ_0$ are the angular momentum operators 
generators of Lorentz transformations centered in $x=0$. Note that in the second integral
in (\ref{generators}), in order to be consistent with the definition of the mapping 
to the flat tangent space, we insert the parallel transported geodesic tangent unit 
vectors $e^{({\lambda})}$ in $x=0$. However, their Taylor expansion does not imply 
corrections to the corresponding integral in eq.~(\ref{expand2}) up to third order 
in $x$, with coefficients of the form $\varpi R$, $R$ being the Riemann tensor. 
A similar argument applies to the integral multiplying the tensor $B$ in eq.~(\ref{expand2}),
as well as to the charged current integral, keeping in mind that $\zeta$ is constant 
at equilibrium.

To summarize, the mean value (\ref{mean}) in a curved spacetime at thermodynamic 
equilibrium can be approximated by an expression which is fully meaningful in the 
familiar Minkowskian quantum field theory
\bea\label{main2}
 && \langle \wO(x) \rangle = \frac{1}{Z} \tr \left( \exp \left[ \zeta \wQ - \beta_\nu(x) \wP^\nu
 + \frac{1}{2} \varpi_{\lambda\nu}(x) \wJ^{\lambda\nu}_x \right. \right. \nonumber \\
 && \left. \left. - \frac{1}{2 \sqrt{\beta^2}} B_{\lambda\sigma\nu}(x) \wTheta^{\lambda\sigma\nu}_x + 
 \ldots \right] \wO(x) \right)_{\rm ren}
\eea
where $\wTheta_x$ are the operators
\be\label{theta}
 \wTheta^{\sigma\lambda\nu}_x \equiv 
 \int \di^3 y \;  (y-x)^\lambda (y-x)^\sigma \wT^{0 \nu}(y) 
\ee
Note that the expression of $\wTheta^{\lambda\sigma\nu}_x$ on the right hand side
of eq.~(\ref{theta}) is neither a local nor a global tensor because it results from 
the integration of a non conserved tensor current. Indeed, it depends on the chosen 
coordinates (the RNC) and, unlike $\wP$ and $\wJ$, on the chosen spacelike hyperplane 
$T\Sigma$. Since the integrand was Taylor expanded over the hypersurface itself, 
the indices of the coordinates $y$ in (\ref{theta}) do not include $\lambda=0$ or 
$\sigma=0$, and so $\wTheta$ is orthogonal to $\beta(x)$ in both the indices $\lambda$ 
and $\sigma$. 

The thermal vorticity $\varpi$ and the tensor $B$ are adimensional in natural units
and are usually very small, so, for most practical purposes all terms in the exponent 
beyond the leading $\wA = -\beta \cdot \wP + \zeta \wQ$ in eq.~(\ref{main2}), denoted
collectively by $\wC$, can be considered as small perturbations and an expansion can 
be done by using Kubo identity
\be\label{kubo}
 \e^{\wA + \wC} \simeq \e^\wA + \int_0^1 \di z \; \e^{z\wA} \, \wC \, \e^{-z\wA} \; 
 \e^\wA
\ee
Thereby, expanding both the exponential operator and its trace $Z$ in eq.~(\ref{main2}), 
$\langle \wO(x) \rangle$ turns out to be the sum of the familiar mean value at 
homogeneous thermodynamical equilibrium with a constant $\beta = \beta(x)$
$$
  \langle \wO(x) \rangle_{\beta(x)} = \frac{\tr \left( \exp \left[ - \beta_\nu(x) \wP^\nu 
  \right] \wO(x) \right)} {\tr \left( \exp \left[ - \beta_\nu(x) \wP^\nu \right] 
  \right)}
$$
and higher order terms involving correlations between $\wO(x)$ and the operators 
$\wJ$ and $\wTheta$, also calculated at the homogeneous thermodynamical equilibrium. 

A physically relevant case is the calculation of the mean value of the stress-energy
tensor, i.e. when $\wO = \wT^{\alpha\gamma}$. In this case, the leading term is the
well known ideal form of the stress-energy tensor
\be\label{ideal}
  \langle \wT^{\alpha\gamma}(x) \rangle_{\beta(x)} = \rho u^\alpha u^\gamma 
  - p \Delta^{\alpha\gamma}
\ee
where $u=\beta/\sqrt{\beta^2}$, $\Delta^{\alpha\gamma} = g^{\alpha\gamma} - u^\alpha 
u^\gamma$ is the projector orthogonal to $u$, $\rho$ and $p$ are the familiar homogeneous 
equilibrium energy density and pressure which are a function of $T,\mu$, or, tantamount, 
$\beta^2,\zeta$ \cite{becalocal}. 

The corrections to theq eq.~(\ref{ideal}) involve the correlators between $\wJ$ and 
$\wT$ and $\wTheta$ and $\wT$. The former were extensively discussed and calculated 
in ref.~\cite{becaquantum} where it was shown that leading corrections are quadratic in 
$\varpi$ and that they are quantum mechanical for a free field, namely they vanish 
in the limit $\hbar \to 0$. They are generally tiny, but tend to become important 
at very low temperature or very large acceleration. In a general curved spacetime 
these corrections are also relevant because, looking at the eq.~(\ref{omega}), $\beta$ 
can have a non-vanishing exterior derivative. This is the case, for instance, for a 
static spherically symmetric gravitational field, where
$$ 
  \beta^\mu = \frac{1}{T_0} (1,0,0,0)
$$
being $T_0$ a constant {\em global} temperature, so that
$$
 T = 1/\sqrt{\beta^2} = \frac{T_0}{\sqrt{g_{00}}}  
$$
which is the Tolman's law, and
$$
  \varpi_{rt} = - \frac{1}{2} \partial_r \beta_t = 
  -\frac{1}{ 2 T_0} \frac{\partial g_{00}}{\partial r}
$$
that is, if $\varpi$ has a longitudinal - along $\beta$ - component which is proportional 
to gravitational acceleration and couples to boosts \cite{becaquantum}.

Let us now turn our attention to the term involving the operator $\wTheta$ in 
eq.~(\ref{main2}). This gives rise to a first-order correction of the mean value 
of the stress-energy tensor, which is linear in the Riemann and Ricci tensors
\be\label{lrt}
 \delta \langle \wT^{\alpha\gamma}(x) \rangle \simeq 
 - \frac{T}{2} B_{\lambda\sigma\nu}(x) \int_0^1 \di z \; 
 \langle \wT^{\alpha\gamma}(x+\ii z\beta) ; \wTheta^{\lambda\sigma\nu}_x \rangle_{\beta(x)}
\ee
where $\langle \wA ; \wB \rangle = \frac{1}{2} \langle \{ \wA, \wB \} \rangle 
- \langle \wA \rangle \langle \wB \rangle$ \footnote{The anticommutator stems from
putting $\wT$ to the right and to the left of the operators on both sides of the
eq.~(\ref{kubo}), then taking the trace and averaging. With this procedure, the 
correlator in eq.~(\ref{lrt}) is manifestly real}. In order to expand the correlator 
in eq.~(\ref{lrt}), the tensors $\wT$ and $\wTheta$ must be decomposed into irreducible 
components with respect to the rotation group of the hyperplane normal to $\beta(x)$. 
Only the components belonging to the same irreducible representation can contract to give
rise to a non-vanishing scalar. The decomposition is well known for the stress-energy 
tensor operator 
\be\label{decomp1}
 \wT^{\alpha\gamma}(x) = \widehat\rho u^\alpha u^\gamma + \widehat\pi
 \Delta^{\alpha\gamma} + \widehat q_\kappa (\Delta^{\alpha\kappa} u^\gamma + 
 \Delta^{\alpha\gamma} u^\kappa) + \widehat\Pi^{\alpha\gamma}
\ee
where $\widehat q$ is orthogonal to $u$ (spin 1 component) and $\widehat \Pi$ is 
orthogonal to $u$ in both indices and traceless (spin 2 component).
The decomposition of the rank 3 tensor requires a little more effort, but eventually 
it does not look much more complicated in view of the transversality of the first 
two indices with respect to the four-velocity
\bea\label{decomp2}
  && \wTheta_x^{\lambda\sigma\nu} = \Delta^{\lambda\sigma} u^\nu \widehat S +
  u^\nu \widehat U^{\lambda\sigma} - (2 \Delta^{\lambda\sigma} \Delta^\nu_\kappa 
  - \Delta^\lambda_\kappa \Delta^{\sigma\nu} \nonumber \\
  && - \Delta^\sigma_\kappa \Delta^{\lambda\nu}) \widehat X^\kappa + \left( 
  u_\tau E^{\tau\lambda\nu\kappa} \widehat Y^{\sigma}_\kappa + u_\tau 
  E^{\tau\sigma\nu\kappa} \widehat Y^{\lambda}_\kappa \right)
  \nonumber \\
  && + \left(\Delta^{\lambda\sigma} \widehat W^\nu + \Delta^{\lambda\nu} \widehat W^\sigma
  + \Delta^{\sigma\nu} \widehat W^\lambda \right) + \widehat Z^{\lambda\sigma\nu}
\eea
where $\widehat S$ is a scalar, $\widehat X$, $\widehat W$ are transverse to $u$ 
(spin 1) and $\widehat U$, $\widehat Y$ are transverse to $u$ in both indices, 
symmetric and traceless (spin 2); $\widehat Z$ is a traceless fully symmetric tensor 
(spin 3) which does not contribute to $T^{\mu\nu}$.
One can now plug the decompositions (\ref{decomp1}) and (\ref{decomp2}) into the
correlator in eq.~(\ref{lrt}) and express it in terms of few scalar coefficients 
depending only on the proper temperature and chemical potential
\bea
 &&\int_0^1 \di z \; \langle \wT^{\alpha\gamma}(x+\ii z\beta) ; \wTheta^{\lambda\sigma\nu}_x 
 \rangle_{\beta(x)} = \chi_\rho u^\alpha u^\gamma \Delta^{\lambda\sigma} u^\nu \nonumber \\
 && + \chi_\pi \Delta^{\alpha\gamma} \Delta^{\lambda\sigma} u^\nu + 
 \chi_v [(2\Delta^{\lambda\sigma} \Delta^{\nu\alpha} - \Delta^{\lambda\alpha}\Delta^{\nu\sigma}
 - \Delta^{\sigma\alpha}\Delta^{\nu\lambda}) u^\gamma \nonumber \\ 
 && + (\alpha \leftrightarrow \gamma)] + \chi^\prime_v [ (\Delta^{\lambda\sigma}\Delta^{\nu\alpha}
 + \Delta^{\lambda\nu}\Delta^{\sigma\alpha} + \Delta^{\sigma\nu}\Delta^{\lambda\alpha}) u^\gamma
  \nonumber \\
 && + (\alpha \leftrightarrow \gamma) ] + \chi_t P^{\alpha\gamma \, \lambda\sigma} u^\nu + 
 \chi^\prime_t (u_\tau E^{\tau\lambda\nu\kappa} P^{\alpha\gamma \, \sigma}_{\quad\;\;\kappa}
 + (\lambda \leftrightarrow \sigma)) \nonumber \\
 \label{lrt3} \\
 && P^{\alpha\gamma}_{\kappa\iota} = \frac{1}{2} (\Delta^\alpha_\kappa 
 \Delta^\gamma_\iota + \Delta^\alpha_\iota \Delta^\gamma_\kappa) -
 \frac{1}{3} \Delta^{\alpha\gamma} \Delta_{\kappa\iota} \label{proje}
\eea
where we have also written the the projector onto the traceless spin 2 component. 

Plugging eq.~(\ref{lrt3}) into eq.~(\ref{lrt}) and using the eq.~(\ref{btensor}), 
the leading correction of $T^{\mu\nu}$ proportional to curvature is obtained
\bea\label{finalt}
 && \delta \langle \wT^{\alpha\gamma}(x) \rangle \simeq \frac{T}{6} 
 (R_{\rho\sigma} \beta^\rho\beta^\sigma + R \beta^2) (\chi_\rho u^\alpha u^\gamma
 + \chi_\pi \Delta^{\alpha\gamma}) \nonumber \\
 && - \frac{T}{6} \chi_t (2 \beta_\rho\beta_\sigma R^{\rho<\alpha \gamma>\sigma} 
   - R^{<\alpha\gamma>} \beta^2)  \nonumber \\
 && + 3 T \chi_v \Delta^{\rho(\alpha} R_{\rho\sigma} \beta^\sigma \beta^{\gamma)}
 + \frac{3}{4} T \chi^\prime_t {\tilde R}^{\rho(\alpha\gamma)\sigma} \beta_\rho\beta_\sigma 
\eea
where $\tilde R^{\mu\nu\rho\sigma} = E^{\mu\nu\lambda\tau}R_{\lambda\tau}^{\;\;\;\rho\sigma}$
is the dual of the Riemann tensor, the round brackets next to indices imply symmetrization
and the angular brackets imply the application of the projector (\ref{proje}). Note that 
the term proportional to $\chi^\prime_v$ has vanishing contraction with $B$ whereas the 
term proportional to $\tilde R$ can be non-vanishing only in a parity and time-reversal 
violating theory. 
The other coefficients $\chi$'s can be calculated with the usual techniques of thermal field
theory by suitably choosing indices in eq.~(\ref{lrt3}) (summing convention does not apply)
\bea
 && \chi_\rho =  \frac{1}{3} \sum_{i=1}^3 
 \int_0^1 \di z \; \langle \wT^{00}(x+\ii z\beta); \wTheta^{ii0} \rangle_{\beta}\\
 && \chi_\pi = \frac{1}{9} \sum_{i,j=1}^3 \int_0^1 \di z \; \langle \wT^{ii}(x+\ii z\beta); 
 \wTheta^{jj0} \rangle_{\beta}\\
 && \chi_v = \frac{1}{6} \sum_{j=1}^3 \int_0^1 \di z \; \langle \wT^{i0}(x+\ii z\beta); 
 (\wTheta^{jji} - \wTheta^{ijj}) \rangle_{\beta} \\
 && \chi_t = \int_0^1 \di z \; \langle \wT^{ij}(x+\ii z\beta); \wTheta^{ij0} \rangle_{\beta} 
 \qquad i \ne j 
\eea
with $\beta=(1/T,0)$. We note that the corrections in eq.~(\ref{finalt}) are certainly
of quantum origin for free fields, that is they vanish in the limit $\hbar \to 0$. The reason 
is that in physical units all curvature related terms $R\beta\beta$, adimensional in 
natural units, must be proportional to $\hbar^2$; it is impossible, by having only $m$ 
and $T$ as scales, to generate classical terms with the dimension of an energy density.  
 
That the mean stress-energy tensor had terms - not related to anomalies - linearly 
proportional to curvature tensors was envisaged in ref.~\cite{baier} and the relevant 
coefficients and the relations between them have been studied by several authors over 
the past few years \cite{roma,batta,tifr,moore,starinets,kaminski}. By comparing with the 
classification in ref.~\cite{baier} we have the identification $-\chi_t/6T = \kappa$ 
and the additional relations $\kappa^* = 0$ as well as $\chi_\pi/6T = \xi_5 = \xi_6$. 
It is still not clear whether the last two additional relations - not predicted in other 
studies - only apply at global equilibrium or if deviations thereof may arise in a 
full calculation without flat spacetime approximation in the calculation of correlation 
integrals. A very important - physical - difference is that in our calculation the 
frame is defined by the four-temperature Killing vector, which, as it seen in eq.~(\ref{finalt}) 
implies non-vanishing corrections to proper energy density and heat flux.  

In summary, we have presented a method to calculate the mean value of the quantum
stress-energy tensor at global thermodynamical equilibrium in a curved spacetime in 
the limit of small curvature, by Taylor expanding the timelike four-temperature 
Killing vector $\beta$ in the density operator. We have shown that the leading 
corrections are proportional to the squared of the exterior derivative of $\beta$
and to the Riemann tensor. This approach clarifies the origin of these terms introduced 
in previous classifications \cite{roma} and provides, in a straightforward fashion,
"Kubo formulae" to calculate relevant coefficients. It would be very interesting to 
to study the phenomenological relevance of these terms in relativistic astrophysics and to 
interpret them in the framework of extended theory of gravity \cite{capozz}.\\
{\em Acknowledgments}. This work was partly supported by Ente Cassa di Risparmio
di Firenze. We are grateful to D. Seminara for useful discussions.

%************************************************************************

%*************************

\end{document}